\documentstyle[11pt,aaspp4]{article}

\def\mathnew{\mathsurround=0pt}
\def\simov#1#2{\lower .5pt\vbox{\baselineskip0pt \lineskip-.5pt
        \ialign{$\mathnew#1\hfil##\hfil$\crcr#2\crcr\sim\crcr}}}
\def\simg{\mathrel{\mathpalette\simov >}}
\def\siml{\mathrel{\mathpalette\simov <}}
\def\Mesz{M\'esz\'aros~}

\received{ 8/31/1996}
\accepted{ 3/20/1997}
\lefthead{\Mesz \& Rees}
\righthead{Poynting Jets from Black Holes and Cosmological GRB}
\slugcomment{Ap.J.(Letters), in press}

\begin{document}

\title{ POYNTING JETS FROM BLACK HOLES\\ AND COSMOLOGICAL GAMMA-RAY BURSTS}

\author{P. \Mesz\altaffilmark{1}}
\affil{Dpt. of Astronomy \& Astrophysics, Pennsylvania State University, 
University Park, PA 16803}
\and
\author{M.J. Rees}
\affil{Institute of Astronomy, Cambridge University, Madingley Road, Cambridge 
CB3 0HA, U.K.}
\altaffiltext{1}{also Center for Gravitational Physics and Geometry, 
Pennsylvania State University}

\begin{abstract}
We discuss the properties of magnetically dominated jet-like outflows from 
stellar mass black holes surrounded by debris tori resulting from neutron star 
disruption. These jets may have narrow cores (along the rotation axis) which 
are almost free of baryons and attain very high bulk Lorentz factors $\simg 
10^6$. The jets give rise to a characteristic MeV to TeV emission as well 
as to relativistic shocks producing the usual MeV bursts. Because the outflow 
is highly directional the properties of the observed gamma-rays would depend on
the viewing angle relative to the rotation axis.  Even for the most intense 
bursts, which under the assumption of isotropic emission and substantial
redshifts would be inferred to emit $10^{52}-10^{53}$ erg, the efficiencies 
required are only $10^{-2}-10^{-4}$.

\end{abstract}

\keywords{gamma-rays: bursts -- black hole physics -- magnetic fields }

\section{ Introduction}

Gamma ray bursts pose two sets of problems. The first is to account for the
required large and sudden energy release. The prime candidate here is the
formation of a compact object or the merger of a compact binary: this can
trigger the requisite energy release (few $10^{51}$ erg s$^{-1}$ for an 
isotropic burst at cosmological distances), with a characteristic dynamical 
timescales  as short as milliseconds. The second problem is how this energy is 
transformed into a relativistically-outflowing plasma able to emit intense 
gamma rays with a nonthermal spectrum. 

The  literature on the second of these problems is already extensive. There
have, in particular, been detailed calculations on the behavior of
relativistic winds and fireballs.  We have ourselves, in earlier papers (e.g. 
Rees \& \Mesz, 1992, 1994, \Mesz, Laguna \& Rees, 1993, \Mesz, Rees \& 
Papathanassiou, 1994 [MRP94], Papathanassiou \& \Mesz, 1996; also 
Paczy\'nski, 1990, Katz, 1994, 
Sari, Narayan \& Piran, 1996) addressed the physical processes in relativistic 
winds and fireballs. Motivating such work is the belief that compact objects 
can indeed generate such outflows.  There  have, however,  been relatively few 
attempts to relate the physics of the outflow to a  realistic model of 
what might energize it. (Among these are early suggestions by Paczy\'nski, 1991,
\Mesz \& Rees, 1992, Narayan, Paczy\'nski \& Piran, 1992, Thompson, 1994, and 
Usov, 1994).  These have involved either the reconversion of a burst of 
neutrino energy into pairs and gamma rays, or else strong magnetic fields. 
Although the former model cannot be ruled out, the beamed neutrino annihilation 
luminosity being marginal in black holes with a disrupted neutron star torus
(Jaroszy\'nski, 1996), initial calculations for neutron star mergers are 
discouraging (e.g. Ruffert et.al. 1997); we here focus attention on magnetic 
mechanisms. In the present paper we try to incorporate our earlier idealized 
models into a more realistic context, and consider some of the distinctive 
consequences of an outflow which is directional rather than isotropic.

\section{ Magnetically Driven Outflows}

Magnetic fields must be exceedingly high in order to  transform rotational 
energy quickly enough into Poynting flux. At the 'base' of the flow, at a 
radius $r_l \sim 10^6-10^7$ cm, where the rotation speeds may be of order $c$, 
the field strength must be at least $\sim 10^{14}$ G to transmit $10^{51}$ ergs 
in a few seconds. These fields are of course higher than those in typical 
pulsars. However , as several authors have noted, the field could be amplified 
by differential rotation, or even by dynamo action (e.g. Thompson \& Duncan, 
1993). If, for instance, a single fast-rotating star collapses or two neutron 
stars spiral together producing (even very transiently) a differentially 
rotating disc-like structure, it need only take a few dynamical timescales for 
the field to amplify to the requisite level (which is, incidentally, at least 
two orders of magnitude below the limit set by the virial theorem).
     
The most severe constraint on any acceptable model for gamma-ray bursts is that 
the baryon density in the outflow must be low enough to 
permit attainment of the requisite high Lorentz factors: the absolute minimum, 
for wind-type models invoking internal dissipation, is $\Gamma \sim 10^2$; 
impulsive models depending on interaction with an external medium require 
$\Gamma \sim 10^3$. Since the overall efficiency is unlikely to exceed 
$10^{-1}$, this requires any entrained baryons to acquire at least $10^3$ 
times their 'pro rata share' of the released energy. There are of course other 
astrophysical situations where an almost baryon-free outflow occurs -- for 
instance the wind from the Crab pulsar, 
which may contain essentially no baryons. However, this is not too 
surprising because the internal dissipation in  pulsars is  far too low to 
generate the Eddington luminosity. On the other hand, in GRBs the  overall 
luminosities are $\simg 10^{13}L_{Ed}$. It is hardly conceivable that 
the fraction channeled into thermal radiation is so low that radiation pressure 
doesn't drive a baryon outflow at some level. The issue is whether this level 
can be low enough to avoid excess 'baryon poisoning'.  

When two neutron stars coalesce, some radiation-driven outflow will be induced 
by tidal dissipation before coalescence (\Mesz \& Rees 1992). When the neutron 
stars have  been disrupted, bulk differential rotation is likely to lead to  
more violent internal dissipation and a stronger radiation-driven outflow.
Almost certainly, therefore, some parts of the outflow must, for some reason, 
be  less accessible to the baryons. 

\section{ Axisymmetric Debris and Jets Around Black Holes}

One such reason might be that the bursts come from a black hole orbited by
a disk annulus or torus (e.g. Paczy\'nski 1991, Levinson \& Eichler, 1993).
This is of course what happens when the
central part of the rotating gaseous configuration has collapsed within
its gravitational horizon; otherwise, there is no reason why  material should 
avoid the centre -- indeed, there is more likely to be a central  density peak 
in any non-collapsed configuration supported largely by rotation (either a
single star or a compact binary after disruption). 

Such a configuration could come about in two ways:
\break
(i) The spinning disk that forms when two neutron stars merge (e.g. Davies 
et.al. 1994), probably exceeds the maximum permitted mass for a single neutron 
star; after viscosity had redistributed its angular momentum, it would evolve 
into a black hole (of 2-3 $M_\odot$) surrounded by a torus of mass 
about 0.1 $M_\odot$ (Ruffert et.al. 1996). 
\break
(ii) The system may result from coalescence of neutron star and black hole
binary. If the hole mass is $\siml 5 M_\odot$, the neutron star would be 
tidally disrupted before being swallowed, leading to a system resembling (i), 
but with a characteristic radius larger by a factor two and a torus mass of
$\sim 1$ instead of $0.1 M_\odot$.  
\break
Numerical simulations yield important insights into the formation of such 
configurations (and the relative masses of the hole and the torus); 
but the Lorentz factors of the outflow are sensitive to much smaller 
mass fractions than they can yet resolve.

It is, however, a general feature of axisymmetric flows around black holes that 
the region near the axis tends to be empty. This is because the hole can swallow
any material with angular momentum below some specific value: within a roughly
paraboloidal 'vortex' region around the symmetry axis (Fishbone \& 
Moncrief, 1976),  infall or outflow are the only options.  Loops of magnetic 
field anchored in the torus can enter this region, owing to 'buoyancy' effects 
operating against the effective gravity due to centrifugal effects at the 
vortex walls, just as they can rise from a flat gravitating disk. These 
would flow out along the axis. There can, in addition, be an ordered poloidal 
field threading the hole, associated with a current ring in the torus. This
ordered field (which would need to be the outcome of dynamo amplification rather
than just differential shearing) can extract energy via the Blandford-Znajek 
(1977) effect. In the latter the role of the torus is mainly to anchor the 
field: the power comes from the hole itself, whose spin energy can amount to 
$\sim 10^{53}$ erg. Irrespective of the detailed field structure, there is good 
reason to expect any magnetically-driven outflow to be less loaded with baryons 
along the rotation axis than in other directions.  Field lines that  thread 
the hole may be completely free of baryons. 

\section{ Baryon-Free Outflows}

As a preliminary, we consider a magnetically-driven outflow in which baryonic
contamination can be neglected.  In the context of many models this may seem an
unphysical limiting case: the difficulty is generally to understand how the
baryonic contamination stays below the requisite threshold.  But the comments 
in the previous section suggest that it is not impossible for the part of the 
flow that emanates from near a black hole and is channelled along directions
aligned with the rotation axis.

There have been earlier discussions (dating back to Phinney, 1982) of
relativistic MHD flows from black holes in AGN.
In  our present context, the values of L/M are larger by at
least ten orders of magnitude. This means that the effects of radiation pressure
and drag are potentially much stronger relative to gravity;  also,  pair
production due to photon-photon encounters is vastly more important. 
For an outflow of magnetic luminosity $L$ and $e^\pm\gamma$ luminosity 
$L_w \siml  L$ channeled into jets of opening angle $\theta$ at a lower radius 
$r_l=10^6 r_6$ cm the initial bulk Lorentz factor is $\Gamma_l \sim L/L_w$, 
and the comoving magnetic field, temperature and pair density are
$B'_l \sim 2.5\times 10^{14} L_{51}^{1/2} \Gamma_l^{-1} r_6^{-1}\theta^{-1}$ G,
$T'_l \sim 2.5\times 10^{10} L_{51}^{1/4} \Gamma_l^{-3/4} r_6^{-1/2} 
\theta^{-1/2}$ K and $n'_l \sim 4\times 10^{32} L_{51}^{3/4} \Gamma_l^{-9/4} 
r_6^{-3/2} \theta^{-3/2}$ cm$^{-3}$ (primed quantities are comoving).
Unless $\Gamma_l >>1$ the jet will be loaded with pairs, very 
opaque and in local thermal equilibrium. It behaves as a relativistic gas 
which is ``frozen in" to the magnetic field, and
expands with $T' \propto r^{-1}$. The lab-frame transverse field 
$B \propto r^{-1}$, and the comoving $B'\sim B/\Gamma$. The comoving 
energy density (predominantly magnetic) is $\epsilon'\propto r^{-2}\Gamma^{-2}$,
and the the pair density is $n' \propto T'^3 \propto r^{-3}$ so $\Gamma \propto 
n'/\epsilon' \propto r$, or $\Gamma \sim \Gamma_l (r/r_l)$.

When the comoving temperature approaches $m_e c^2/k$ the pairs start to 
annihilate and their density drops exponentially, but as long as the scattering 
depth $\tau'_T > 1$ the annihilation photons remain trapped and continue to 
provide inertia, so $T' \propto r^{-1}$, $\Gamma \propto r$ persists until 
$T'_a \sim 0.04 m_ec^2 \simeq$ 17 keV at a radius $r_a$, where 
$\tau'_T (r_a) \sim 1$. Between $r_l$ and $r_a$ this leads to $(r_a/r_l) \sim 
(T'_l/T'_a) \simeq 10^2 L_{51}^{1/4} \Gamma_l^{-3/4} r_6^{-1/2} \theta^{-1/2}$, 
with $\Gamma_a \sim \Gamma_l (r_a/r_l) \simeq 10^2 L_{51}^{1/4} \Gamma_l^{1/4} 
r_6^{-1/2} \theta^{-1/2}$. At $r_a$ the adiabatic density $n'_{a,ad} \sim n'_l 
(r_a/r_l)^3 \simeq 4\times 10^{26}$ cm$^{-3}$ is mostly photons, while the 
pair density from Saha's law is $n'_a \sim 1.5 \times 10^{18} \Gamma_l 
r_6^{-1}$, the photon to pair ratio being $\sim 10^8$. The annihilation photons 
streaming out from $r_a$ appear, in the source frame, as photons of energy 
around $\Gamma_a 3 k T'_a  \sim 5 L_{51}^{1/4} \Gamma_l^{1/4} r_6^{-1/2} 
\theta^{-1/2}$ MeV.

Beyond $r_a$ the lab-frame pressure $B^2 \propto r^{-2}$ but 
the inertia is drastically reduced, being provided only by the surviving 
pairs, which are much fewer than the free-streaming photons.
In the absence of any restraining force, $\Gamma \propto n'/\epsilon'
\propto n'/B'^2 \propto B^2/n'$, and the gas would accelerate much faster than 
the previous $\Gamma \propto r$. (The pair density is still above that needed 
to carry the currents associated with the field, analogous to the 
Goldreich-Julian density, so MHD remains valid). However, the Compton drag 
time remains very short, since even after $\tau'_T < 1$ when most photons 
are free-streaming, the pairs experience multiple scatterings with a small 
fraction of the (much more numerous) photons for some distance beyond $r_a$.

One can define an ``isotropic" frame moving at $\Gamma_i \propto r$, in which 
the annihilation photons are isotropic. In the absence of the magnetic pressure,
the drag would cause the electrons to continue to move with the radiation at 
$\Gamma_i \propto r$. The magnetic pressure, however, acting against a much 
reduced inertia, will tend to accelerate the electrons faster than this, and 
as soon as $\Gamma \simg \Gamma_i$, aberration causes the photons to be 
blueshifted and incident on the jet from the forward direction, so the
drag acts now as a brake. In the isotropic frame the jet electron energy is
$\gamma=\Gamma/\Gamma_i$ and its drag timescale is that needed for it to 
encounter a number of photons whose cumulative energy after scattering equals 
the energy per electron, $t_{dr,i} \sim m_e c^2/(u_{ph,i}\sigma c \gamma) = 
(m_e c^2 4\pi r^2 \Gamma_i^3 / L \sigma_T \Gamma)$. In the lab frame this is 
$\Gamma_i$ times longer, and the ratio of the drag time to the expansion time 
$r/c$ must equal the ratio of the kinetic flux to the Poynting flux, $n'_j 
m_ec^2 \Gamma^2 / [(B_l^2/4\pi)(r_l/r)^2]$, where $\sigma$ is the scattering 
cross section and $n'_j$ is the comoving pair density in the jet. This 
is satisfied for $\Gamma \sim \Gamma_a$ at $r_a$, and since the drag time is 
much shorter than the annihilation time, the pair number is approximately 
frozen, and $\Gamma \propto r^{5/2}$ for $r > r_a$.  The upscattered photons 
will, in the observer frame, appear as a power law extension of the 
annihilation spectrum, with photon number index -2, extending from 
$\sim 0.12 \Gamma_a m_ec^2$ to $\siml \Gamma_j m_ec^2$.

The acceleration $\Gamma\propto r^{5/2}$ abates when the annihilation 
photons, of isotropic frame energy $0.12 m_ec^2(r_a/r)$, are 
blueshifted in the jet frame to energies $\simg m_ec^2$. Their directions are 
randomized by scattering, and collisions at large angles above threshold lead 
to runaway $\gamma\gamma \to e^\pm$ (the compactness parameter still being 
large). This occurs when $\Gamma_p \sim 10^7 L_{51}^{1/4} \Gamma_l^{1/4} 
r_6^{-1/2} \theta^{-1/2}$, at $r_p \sim 10^{10} L_{51}^{1/4} \Gamma_l^{-3/4} 
r_6^{1/2} \theta^{-1/2}$ cm. Thereafter the threshold condition implies
$\Gamma \propto r^2$, until the inertial limit $\Gamma_{in}\sim 10^9 
L_{51}^{1/4} \Gamma_l^{1/4} r_6^{1/2} \theta^{-1/2}$ is reached at $r_{in}\sim 
10^{11}  L_{51}^{1/4} \Gamma_l^{-3/4} r_6 \theta^{-1/2}$ cm. Besides going into
pairs, a reduced fraction of the Poynting energy may continue going into a 
scattered spectrum of number slope -2 up to $\Gamma_{in} m_ec^2$.

However, in an outflow of $L\sim 10^{51} L_{51}$ erg s$^{-1}$ maintained for
a time $t_w\sim$ few seconds, the relativistic jet would have to 
push its way out through an external medium, with consequent dissipation at 
the front end, as in double radio sources. Except for a brief initial transient 
($\ll 1$ s in observer frame) the shock will be far outside the characteristic 
radii discussed so far. The external shock will be ultrarelativistic, and 
slows down as $r^{-1/2}$. The jet material, moving with $\Gamma \gg 1$,
therefore passes through a (reverse) shock, inside the contact 
discontinuity, which strengthens as $r^{1/2}$ as the external shock slows down. 
Since it is highly relativistic and the medium highly magnetized, this reverse 
shock can emit much of the overall burst energy on a time $\sim t_w$. 
When $t\sim t_w$ the external shock has reached $r_d \sim 10^{16} 
L_{51}^{1/4} n_o^{-1/4} t_w^{1/2} \theta^{-1/2}$ cm, where $n_o$ is the 
external density, and the Lorentz factor has dropped to $\sim 10^3 
L_{51}^{1/8}n_o^{-1/8} t_w^{-1/4} \theta^{-1/4}$. This, as well as the
expected radiation of the shocks at (and after) this stage is rather
insensitive to whether the initial Lorentz factor is indeed $\sim 10^6-10^8$, 
or whether baryon loading has reduced it to $\sim 10^3$. After this the 
flow is in the impulsive regime and produces an ``external" shock GRB on an 
observer timescale $r_d/c\Gamma^2 \simeq t_w$, as in, e.g. \Mesz \& Rees, 
1993, MRP 1994. 

\section{Radiation from High $\Gamma$ Magnetically Driven Jets}

If the jet is indeed baryon-free, and therefore has a Lorentz factor $\sim
10^7-10^9$, an extra mechanism can tap the Poynting energy along its length 
(before it runs into external matter) -- namely, interaction of pairs in the 
jet and annihilation photons along the jet with an ambient radiation field. 
In our context, this would be radiation emitted by the torus or 
baryon-loaded wind that is expected to flow outward, mainly in directions away 
from the rotation axis, forming a funnel that surrounds the jet.  

The ambient radiation causes an additional drag, which limits the 
terminal Lorentz factor of the jet below the values calculated in \S 4 in 
those regions where this is important. As a corollary, the Poynting 
flux is converted into directed high energy radiation, after pair creation in 
the jet by interaction with the annihilation radiation.  
We cannot generally assume that the ambient radiation field is uniform across 
the whole jet, because it may not be able to penetrate to the axis, but the 
boost in photon energy can in principle be $\siml \Gamma^2$. This is similar 
to what is discussed in AGN but with more extreme Lorentz factors and radiation 
densities. Since the ambient radiation has a luminosity $\simg L_{Ed}$, and the 
burst luminosity is larger by $10^{12}-10^{13}$, this mechanism is significant 
only when the jet Lorentz factor has the very high values $\simg 10^6$ 
characteristic of these baryon-free outflows. 

The photons from the sides of the funnel emitted into the jet have 
energies $x_f=E_\gamma /m_ec^2\sim 1/20$.
When they pass transversely into the jet, and are scattered by 
``cold" pairs, the scattering will be in the K-N regime for any $\Gamma >20$. 
The electron will therefore recoil relativistically, acquiring a Lorentz factor 
$\sim \Gamma/20$ (in the comoving frame), and will then cool by synchrotron 
emission, isotropic in the jet frame. This radiation will, in the observer 
frame, be beamed  along the jet and blueshifted by $\Gamma$. One readily sees 
that the net amplification (energy/incident energy) is $\sim \Gamma^2$.
If the external photons instead interact with one of the beamed gamma-rays of
energy $E$ in the source frame (typically near threshold for pair production) 
the resultant pair will have a Lorentz factor of order  $\Gamma / (E/m_e c2)$ 
in the comoving frame, and will again emit synchrotron, yielding the same 
amplification factor as before. 

The interaction of the ambient photons with the beamed annihilation
photons in the jet leads to a complicated cascade, where the jet Lorentz
factor must be calculated self-consistently with the drag exerted by the
ambient photons.  A schematic cascade at $r\sim r_p$ where $\Gamma_p \sim 
10^7$, would be as follows. \break
A) Interactions near $r_p$ between funnel photons of lab frame energy $x_f 
\sim 10^{-1}$ and beamed annihilation jet photons of lab frame energy $x_1 
\sim x_a \sim 10$ (produced by pair recombination in the outflow) lead to 
$e^\pm$ of energy $\gamma_2 \sim x_1 \sim 10$ (lab), or $\gamma'_2 \sim 
\Gamma_p/ \gamma_2 \sim 10^6$ (comoving jet frame). In the comoving magnetic 
field $B' \sim 6\times 10^3$ G these pairs produce synchrotron photons of 
energy $x'_2 \sim 10^1$ (comoving), or $x_2 \sim x'_2 \Gamma_p 2\times 10^8$
(lab). \break
B) Photons $x_2$ interact with ambient photons $x_f$ to produce pairs of energy 
$\gamma_3 \sim x_2 \sim 2\times 10^8$ (lab), or $\gamma'_3 \sim \gamma_3 / 
\Gamma_p \sim 2\times 10^1$ (comoving). In the same comoving field 
these produce synchrotron photons $x'_3 \sim 10^{-8}$ (comoving) or $x_3\sim 
10^{-1}$ (lab). \break
C) Photons $x_3 \sim 10^{-1}$ are below threshold for pair production with 
funnel photons $x_f\sim 10^{-1}$, ending the cascade. The resulting photons 
have lab energes $E_3 \sim x_3 m_ec^2 \sim 50 L_{51}^{1/4} \Gamma_l^{1/4} 
r_6^{-1/2}\theta^{-1/2}$ KeV. 
Of course, in a more realistic calculation the self-consistent jet Lorentz 
factor may vary across the jet, and one needs to integrate over height. 
However this simple example illustrates the types of processes involved.

\section{ Discussion}

>From the scenario above it follows that Poynting dominated (or magneticaly 
driven) outflows from collapsed stellar objects would lead, if viewed along the 
baryon-free jet core, to a $\gamma$-ray component resembling the GRB from the 
external shocks in 'impulsive fireball' models. However the reverse shock is 
here relativistic, and may play a larger role than in impulsive models. 
Viewed at larger angles, the GRB component would ressemble that from internal 
shocks (see below). The characteristic duration $t_w$ and the (possibly 
complicated) light curve are controlled by the details of the Poynting 
luminosity production as a function of time, e.g. the cooling or accretion time 
of the debris around a centrally formed black hole.

Besides the 'standard' GRB emission, Poynting dominated outflows viewed along 
the jet core would be characterized by additional radiation components from the 
jet itself (\S\S 4, 5). The annihilation component peaks at $E_1 \sim 5 
L_{51}^{1/4} \Gamma_l^{1/2} r_6^{-1/2} \theta^{-1/2}$ MeV (or 50 MeV if
$\theta \sim 0.1$, $\Gamma_l\sim 10$), with a power law extension of photon 
number slope -2 going out to $\siml \Gamma_p m_ec^2 \sim 5 L_{51}^{1/4} 
\Gamma_l^{1/2} r_6^{-1/2} \theta^{-1/2}$ TeV, if ambient photon drag limits 
$\Gamma$ to $\siml \Gamma_p$ (otherwise, it could extend to $\Gamma_{in} m_ec^2 
\sim 500$ TeV).  If, as argued by Illarionov \& Krolik (1996) for AGN, an outer 
skin of optical depth unity protects the core of the jet from ambient photons, 
this annihilation component could have a luminosity not far below the GRB MeV 
emission, and the $\Gamma$ of the jet core would not be limited by ambient 
drag, only that of the skin. However, the skin depth is hard to estimate 
without taking the drag self-consistently into account. The cascade from the 
interaction of ambient (funnel) photons with jet photons leads to another jet 
radiation component. The simplified estimate of \S 5 gives $E_3 \sim 50 
L_{51}^{1/4} \Gamma_l^{1/4} r_6^{-1/2}\theta^{-1/2}$ KeV as a characteristic 
energy, with a power law extension. The amplification factor of the cascade is 
$A_c \siml \Gamma^2 \siml 10^{14} L_{51}^{1/2} \Gamma_l^{1/2} r_6^{-1} 
\theta^{-1}$, and since the funnel emits $\sim L_{Edd}$, the luminosity could 
be a (possibly small) fraction of $L$. The duration of these components 
is $\sim t_w$, preceding the normal MeV burst by $\sim t_w$, and it may be 
more narrowly beamed, arising from regions with $\Gamma_p \sim 10^7$, as 
opposed to $\Gamma \siml 10^3$ for the MeV components, and $\Gamma_a \sim 10^2$ 
for the peak annihilation component.

Gamma-rays will be detected for a time $t_w$ (longer if external shocks are
expected). There may also be prolonged after effects, e.g. \Mesz \& Rees, 1997,
with X-ray and optical fluxes decreasing as a power law in time.  In a rotating 
black hole - torus system with a super-Eddington outflow the baryon 
contamination will be minimal along the rotation axis and will increase at 
larger angles to it. A narrow, largely baryon-free jet core such as discussed 
in \S\S 4 and 5, would be surrounded by a debris torus producing a 
baryon-loaded, slower outflow acting as a funnel, which injects X-ray seed 
photons into the jet. This slower outflow would carry a fraction of the GRB 
luminosity $L$ in kinetic energy form. For a slow outflow $\Gamma
\sim 10$, this kinetic energy can be reconverted into nonthermal X-rays after 
a time $t_x \sim$ day if it is shock-heated at a radius $r\sim 10^{15}$ cm,
either by Alfv\'en wave heating, or by interaction with a pre-ejected 
subrelativistic shell of matter of $\sim 10^{-3} M_\odot$ as has been detected
in SN 1987a. This could lead to the substantial X-ray afterglow detected $\sim$
days later in GRB 970228 (Costa, et al., 1997).

The observed time structure and luminosity of the gamma-ray burst would depend 
on the angle between the rotation axis and the line of sight. Viewed obliquely, 
the outflow has $\Gamma \siml 10$ and only an X-ray transient with some 
accompanying optical emission would be seen. At smaller angles to the jet axis, 
outflows with $\Gamma \siml 10^2$ would also be seen (which might originate at 
$r > 10^{10}$ cm by entrainment of slower baryonic matter by the faster jet, 
or they might already originate closer in), and the predominant radiation 
would arise from internal shocks (Rees \& \Mesz, 1994, Papathanassiou \& \Mesz,
1996), which can have complex, multiple-peaked light curves. Closer to 
the rotation axis, outflows with $\Gamma\sim 10^3$ may dominate, either at large
radii or already lower down, and radiation from external (deceleration) shocks 
would be prominent (e.g. MRP94). Nearest to the rotation axis, if there is 
indeed an almost baryon-free core to the jet, the annihilation radiation
power law component, the relativistic reverse shock radiation and the cascade 
process (\S\S 4,5) yield significant extra contributions, which arrive ahead 
of the external shock burst.  The luminosity function  for the burst population 
may in part (or even entirely) be determined by the beam angle distribution, 
jet edge effects and the observer angle relative to the jet axis.  A time 
variability would be expected from intermittency in the Poynting flux 
extraction, or wobbling of the inner torus or black hole rotation axis, on 
timescales down to $t_v \sim r_l/c \sim 10^{-3}- 10^{-4}$ s. 

If the Poynting flux is derived from the accretion energy of a NS-NS remnant 
torus of 0.1 $M_\odot$, and the gamma-rays are concentrated within an 
angle $\theta \sim 10^o$, the efficiency of conversion of rest mass 
into magnetic energy need only be $10^{-4}$ to generate a burst which (if 
isotropic) would be inferred to have an energy of $10^{51}$ ergs, and only 
$10^{-2}$ to simulate a $10^{53}$ erg isotropic burst.  For a BH-NS merger the 
torus may be $\sim 1 M_\odot$, and the corresponding magnetic efficiencies 
required are $10^{-5}$ and $10^{-3}$ respectively. Thus, even bursts whose 
inferred isotropic luminosites are $10^{52}-10^{53}$ erg, either due to high 
redshifts or exceptional observed fluxes, require only very modest efficiencies.
Further reductions in the required efficiencies are possible if the Poynting 
flux is due to the Blandford-Znajek mechanism, which can extract $\siml 
10^{-1}$ of a near-maximally rotating Kerr black hole rest mass (as would be 
expected from a NS-NS merger), leading to equivalent isotropic energies 
$\sim 10^{53}(4\pi/2\pi\theta^2) \gg 10^{53}$ erg. Even without the latter, 
somewhat speculative, possibility, it is clear that jet-like Poynting flows 
from tori around black holes can produce even the most
intense bursts with comfortably low efficiencies. The detailed burst properties 
-- particularly the rapid variability and bursts seen along the jet core, the 
blueshifted annihilation and high energy cascade photons-- can help to pin 
down the parameters of the model, while the afterglow can provide information 
on the dynamics and mass flux in directions away from the jet axis.

\acknowledgements
We thank NASA NAG5-2857, NATO CRG-931446 and the Royal Society for support, 
the Institute for Advanced Studies, Princeton, for its hospitality, and 
Ira Wasserman for useful comments.


\begin{thebibliography}{}

\bibitem[Blandford \& Znajek, 1977]{bd77} Blandford, R.D. \& Znajek, R., 1977,
 \mnras, 179, 433
\bibitem[Costa, et al., 1997]{co97} Costa, E., et al., 1997, IAU Circ. 6576
\bibitem[Davis, et.al. 1994]{da94} Davis M.B., Benz W., Piran T. \& 
 Thielemann F.K., 1994, \apj, 431, 742
\bibitem[Fishbone \& Moncrief, 1976]{fm76} Fishbone, L.G. \& Moncrief, V.,
 1976, \apj, 207, 962
\bibitem[Jaroszy\'nski, 1996]{ja96} Jaroszy\'ski, M., 1996, Astron.Ap., 305, 839
\bibitem[Katz, 1994]{ka94} Katz, J.I., 1994, \apj, 422, 248
\bibitem[Krolik \& Illarionov, 1996]{ki96} Krolik, J. \& Illarionov, A., 1996,
 \apj, in press
\bibitem[Levinson \& Eichler, 1993]{le93} Levinson, A \& Eichler, D., 1993, 
 \apj, 418, 386
\bibitem[M\'esz\'aros \& Rees, 1992]{mr92} M\'esz\'aros, P \& Rees, M.J.,
  1992, \apj, 397, 570
\bibitem[M\'esz\'aros \& Rees, 1993]{mr93} M\'esz\'aros, P \& Rees, M.J.,
  1993, \apj, 405, 278
\bibitem[M\'esz\'aros, Rees \& Papathanassiou, 1994]{mrp94} M\'esz\'aros, P,
  Rees, M.J. \& Papathanassiou, H., 1994 [MRP94], \apj, 432, 181.
\bibitem[M\'esz\'aros \& Rees, 1997]{mr97} M\'esz\'aros, P \& Rees, M.J.,
 1997, \apj, 476, 232
\bibitem[Narayan, et.al. 1992]{napapi92} Narayan, R., Paczy\'nski, B. \& 
 Piran, T. 1992, \apjl, 395, L83.
\bibitem[Paczy\'nski, 1990]{pa90} Paczy\'nski, B., 1990, \apj, 363, 218
\bibitem[Paczy\'nski, 1991]{pa91} Paczy\'nski, B., 1991, Acta.Astron. 41, 257
\bibitem[Papathanassiou \& M\'esz\'aros, 1996]{pm96} Papathanassiou, H. \&
 M\'esz\'aros, P., \apjl, 471, L91
\bibitem[Phinney, 1982]{ph82} Phinney, S., 1982, Ph.D. thesis, Cambridge 
 University
\bibitem[Rees \& M\'esz\'aros, 1992]{rm92} Rees, M.J. \& M\'esz\'aros, P,
 1992, \mnras, 258, P41.
\bibitem[Rees \& M\'esz\'aros, 1994]{rm94} Rees, M.J. \& M\'esz\'aros, P,
 1994, \apjl, 430, L93.
\bibitem[Ruffert, et.al. 1996]{ru96} Ruffert, M., Jahnka, H-T, Takahashi K. \&
 Schaefer G. 1996, A\&A, in press
\bibitem[Sari, Narayan \& Piran, 1996]{sanapi96} Sari, R, Narayan, R. \&
 Piran, T., 1996, \apj, in press
\bibitem[Thompson \& Duncan, 1993]{td93} Thompson, C. \& Duncan, R.C., 1993,
 \apj, 408, 194
\bibitem[Thompson, 1994]{thom94} Thompson, C., 1994, \mnras, 270, 480
\bibitem[Usov, 1994]{us94} Usov, V.V., 1994, \mnras, 267, 1035


\end{thebibliography}
\end{document}